\documentclass[journal]{IEEEtran}
\makeatletter
\def\ps@headings{%
\def\@oddhead{\mbox{}\scriptsize\rightmark \hfil \thepage}%
\def\@evenhead{\scriptsize\thepage \hfil \leftmark\mbox{}}%
\def\@oddfoot{}%
\def\@evenfoot{}}
\makeatother
\usepackage{tabularx}
\usepackage{multirow}
\usepackage{booktabs}
\usepackage[utf8x]{inputenc}
\usepackage{framed, color}
\usepackage{mathtools}
\usepackage[super]{nth}
\usepackage{url}
\usepackage{subfig}
\usepackage{hyphenat}
\usepackage{ifthen}
\usepackage{tikz}
\usepackage[square, numbers, sort&compress]{natbib}
\usetikzlibrary{positioning,calc}
\usepackage{epstopdf}
\usepackage{nameref}
\usepackage{chngcntr}
\usepackage{amsmath}
\usepackage{graphicx}
\usepackage{caption}
\usepackage{adjustbox}
\usepackage{soul}
\usepackage[T1]{fontenc}
\usepackage{authblk}
\usepackage{changes}
\setremarkmarkup{\textcolor{red}{\bf{[#2]}}}
\counterwithin*{paragraph}{subsubsection} % removes paragraph from the subsubsections
\counterwithin*{paragraph}{section}
\def\remove#1{}

\begin{document}

%\input{abstract}
 %\title{An experiment in distributed Internet address management using blockchains}

% \makeatletter
% \def\ps@pprintTitle{%
% 	\let\@oddhead\@empty
% 	\let\@evenhead\@empty
% 	\let\@oddfoot\@empty
% 	\let\@evenfoot\@oddfoot
% }
% \makeatother	
 %\corref{mycorrespondingauthor}

\title{An experiment in distributed Internet address management using blockchains}
 \author[1]{Stefano Angieri}
 \author[1]{Alberto  Garc\'{i}a-Mart\'{i}nez}
 \author[2]{Bingyang Liu\thanks{Corresponding Author: Bingyang Liu, email: liubingyang@huawei.com}}
 \author[3]{Zhiwei Yan}
 \author[2]{Chuang Wang}
 \author[1]{Marcelo Bagnulo}
 \affil[1]{Universidad Carlos III de Madrid}
 \affil[2]{Huawei}
 \affil[3]{China Internet Network Information Center}
 \renewcommand\Authands{ and }

% \author{
 %	Stefano Angieri\\
 	%\texttt{UC3M}\\
 %	\and
 %	Alberto  Garc\'{i}a-Mart\'{i}nez\\
 %	\texttt{first2.last2@xxxxx.com}
%	Marcelo Bagnulo\\
%	\texttt{first2.last2@xxxxx.com}
%	 }
% \title{An experiment in distributed Internet address management using blockchains}	
 
 %\title{An experiment in distributed Internet address management using blockchains
 %}

% \author{
 %    Stefano Angieri, Alberto Garc\'{i}a-Mart\'{i}nez, Marcelo Bagnulo, Bingyang Liu, Chuang Wang, Zhiwei Yan
  %   \newline Corresponding Author: Bingyang email: stocazzo
   %}

%%\author{}

%\author[a]{Stefano Angieri}
%\author[a]{Alberto Garc\'{i}a-Mart\'{i}nez} 
%\author[a]{Marcelo Bagnulo}
%\author[b]{Bingyang Liu}
%\author[b]{Chuang Wang}
%\author[c]{Zhiwei Yan}

\maketitle
%\address[a]{Universidad Carlos III de Mardid}
%\address[b]{Huawei}
%\address[c]{China Internet Network Information Center}
 
%\begin{document}

		%%\author{Elsevier\fnref{myfootnote}}
		%%//\address{Radarweg 29, Amsterdam}
		%%\fntext[myfootnote]{Since 1880.}
		
	\begin{abstract}
	The current system to manage the global pool of IP addresses is centralized in five transnational organizations, the Regional Internet Registries (RIRs). Each of these RIRs manage the address pool for a large number of countries. Because the RIRs are private organizations, they are subject to the legal framework of the country where they are based. This configuration results in a jurisdictional overflow from the legal framework of the countries where the RIR is based to all the countries that the RIRs are serving (the countries served by the RIRs de facto become subjects of the legal system of the country where the RIR is hosted). The situation is aggravated by the deployment of new security techniques such as the Resource Public Key Infrastructure (RPKI) and BGP security (BGPsec), that enable enforcement of allocations by the RIRs.\newline \indent
	In this paper we present InBlock, a blockchain-based distributed governance body aimed to provide de-centralized management of IP addresses. InBlock also aims to fulfil the same objectives as the current IP address allocation system, namely, uniqueness, fairness, conservation, aggregation, registration and minimized overhead. InBlock is implemented as a Decentralized Autonomous Organization, i.e., as a set of blockchain's smart contracts in Ethereum. Any entity may request an allocation of addresses to the InBlock registry by solely performing a (crypto)currency transfer to the InBlock. The fee required, along with the annual renewal fee, serves as a mechanism to deter stockpiling and other wasteful practices. \newline \indent
	As with any novel technology, there are many open questions about the usage of blockchains to build an IP address registry. For this reason, we believe that practical experimentation is required in order to have hands-on experiences about such a system. We propose to conduct an experiment on distributed address management using InBlock as a starting point to inform future directions in this area.
	
\end{abstract}

%\begin{keyword}

%	IPv6 address registry \\
%	Blockchain \\
%	Governance

%	\end{keyword}

%\end{frontmatter}

%	\linenumbers

\section{Introduction}
Internet governance is loosely defined as the development and application of the set of norms and rules that shape the use of the Internet \cite{IGdef}. As the complex system the Internet has become, Internet governance is performed at different dimensions and by different entities. For example, the set of rules that are used by the systems to communicate, i.e., the so-called protocols, are defined by institutions with a technical focus such as the Internet Engineering Task Force (IETF). Some policy issues are regulated by national governments through legal regulation, such as those related with content blocking, privacy protection, fundamental human rights, etc. In this paper, we focus on the administration of the Internet Protocol (IP) addresses used to identify and locate of the end points of Internet communications.\newline \indent
IP addresses are globally administered by the Internet Corporation for Assigned Names and Numbers (ICANN), a private non-profit organization. The global pool of IP addresses is managed through a hierarchical structure rooted in ICANN. ICANN delegates ranges of IP addresses to five Regional Internet Registries (RIRs), the second tier of the hierarchy. RIRs allocate address blocks to Local Internet Registries (LIRs, usually Internet Service Providers). End-users normally obtain addresses from the LIRs and in some cases, directly from the RIRs. \newline \indent
While this arrangement has generally been successful, it comes with some rough edges. We argue that this structure for the governance of the global pool of IP addresses results in a \textit{jurisdictional overflow} of the countries where ICANN and the RIRs are hosted into all the other countries served by ICANN and the RIRs. ICANN and the RIRs are private organizations that operate within the legal framework of the countries where they are based. However, they each manage the IP addresses for a large set of countries. This implies that for most countries in the world, the management of a critical Internet resource such as the IP addresses operates under the legal framework of a foreign country. In addition, for these countries, any legal action involving IP address management will be settled in a foreign court of law, making these resources \textit{de facto} subject to laws of a foreign state, as observed in \cite{HZ04}. These concerns have been recently exacerbated by the development of new Internet security techniques, as we describe next.\newline \indent
Over the last few decades, the Internet has become part of the critical infrastructure for most countries. The increasing concern to guarantee its availability has resulted in the design, deployment and adoption of new security tools, such as the Resource Public Key Infrastructure (RPKI) and the security extension to the BGP routing protocol (BGPsec). These tools aim to provide cryptographic guarantees that whoever is claiming to have an Internet addressing resource is indeed the legitimate holder of the resource according to the defined allocation rules, preventing prefix hijacking attacks and other vulnerabilities. As such, these mechanisms provide the entities in the hierarchy of the allocation system (i.e., RIRs, NIRs and LIRs) a capability that they lacked so far, namely the capacity to actually enforce the allocations in real time. In particular, they allow entities in the allocation hierarchy to arbitrarily override an existing IP allocation \cite{HCRG}. So, if/when these cryptographic techniques are widely adopted, the Internet Registries will be able to invalidate allocations, disconnecting whole networks from the Internet, if so dictated by their governing bodies. We note again that a mismatch exists between the geographical scope in which legal, operational or even political decisions are taken (the countries where the RIRs are based) and their effects (the whole world). This situation has raised a number of concerns and it may be one of the reasons behind the lag in the adoption of such technologies. Note that the attacks they are designed to prevent are very real, so security measures to protect the Internet are indeed needed. \newline \indent
The current hierarchical design for managing Internet addresses was probably the most natural one when it was created. IP addresses come from a single global pool and in order to properly perform its function, global uniqueness must be guaranteed, i.e., the system must prevent that the same address is simultaneously allocated to two different parties. When the Internet was designed, the straightforward way of accomplishing this was to rely on a hierarchical structure to manage the allocations of IP addresses, preventing the allocation of the same resource twice.\newline \indent
However, new opportunities for managing namespaces (in our case, the IP address namespace) surface with the recent introduction of the blockchain technology. Blockchains are distributed databases that are controlled through consensus. By design, they prevent any central authority to control the content of the database. The blockchain technology provides an opportunity to explore alternative Internet governance approaches. In particular, it provides the grounds to create a blockchain-based registry for Internet resources that is not controlled by any single entity. This is appealing, as it would more accurately reflect the current Internet reality as a global network. \newline \indent
In this paper, we present InBlock, a distributed autonomous Internet address registry. The proposed approach uses blockchain technology to perform the IPv6 address registry functions. While InBlock supports both IPv4 and IPv6, we focus on IPv6, as IPv6 has a large remaining pool of unassigned addresses, while the vast majority of the IPv4 address space has already been assigned \cite{Levin14}. InBlock makes different trade-offs than the current hierarchical allocation system. First and most importantly, InBlock is not centrally controlled, but, as any blockchain-based mechanism, it is controlled through distributed consensus. InBlock provides a distributed, automatic, irrevocable, tamper-free, publicly accessible, privacy-preserving resource allocation mechanism, designed with the appropriate (economic) incentives to enforce address conservation. In addition, this solution is compatible with the cryptographic security architecture developed for the Internet routing system.\newline \indent 
InBlock is a set of programmes that autonomously run in the blockchain, performing the functions of an IP address registry, as defined by \cite{RFC2050}. In a nutshell, InBlock has a block of globally routable IP addresses to allocate. To request an address allocation, an entity transfers a fee to InBlock. Once the fee transaction has been verified, InBlock annotates in the blockchain the allocation of a prefix to the requester, serving as a ledger of the address assignment for any interested party accessing to the blockchain. Note that, following the paradigm of \textit{code is law}, the governance bylaws are defined in the InBlock code and executed in the same way by any node validating the blockchain. The human action is limited to the initial definition of the allocations rules. As a blockchain-based registry, it is not controlled by any single entity, so when applied to IP address assignment, it results in a mechanism that reflects more accurately the current Internet reality as a global network. This approach provides clear and transparent rules without any kind of human discretion.  Address conservation is preserved by a fee mechanism which mimics the current fees charged by the RIRs to the recipients of IP address allocations. In addition, InBlock provides the means to become the authoritative database in which the assignee of the prefix can associate the basic information for the Internet routing system to operate, currently stored in the Internet Routing Registries \cite{KM17}, along with the cryptographic information that may be used to secure the routing system. \newline \indent
It is worth to note that InBlock is designed as an additional IPv6 registry and not as a replacement for the current IANA/RIR based one. The current hierarchical system has stood the test of time, and the ICANN plus RIR system provides services that are appreciated by the Internet community. Therefore, we do not devise a migration strategy in which the information currently hold by the RIRs is transferred to a blockchain based mechanism. In addition, we are specifically not engaged in fostering a competition of InBlock with the current allocation system, for example, by exploiting the lower costs that an automatic address assigning mechanism would exhibit in comparison with an organization-based one. We envision InBlock as just an alternative for entities deeply concerned about the mismatch in jurisdictions, including the RIRs themselves, which may not want to be responsible for address allocations that are subject to legal processes either in their jurisdiction of in the area in which they operate. \newline \indent
As it is the case for most new disrupting technologies, there are many open questions about blockchains including how sustainable are they, how they will evolve, among many others. We now lie on a crossroad: we have a technology that has the potential to bring decentralized governance to the Internet, but it is still too immature to be fully adopted. On one hand it is in the Internet’s genetic code to embrace innovation, but on the other hand, there is too much uncertainty about blockchains and too much at stake to simply adopt an Internet-wide Blockchain-based registry.  \newline \indent
For the aforementioned reasons, we believe it would be beneficial for the Internet community to perform a series of experiments on decentralized Internet governance using the blockchain technology. The proposal is to allocate a small IP address block out of the global address space for the experiment and to create one or more blockchain-based registries to manage the allocations out of that address block, with a predefined lifetime. Such an experiment would enlighten the community with a hands-on experience to inform future directions regarding decentralized Internet governance.\newline \indent
The rest of the paper is structured as follows: 
In section \ref{sec:BackgroundIp} we describe the current IP address allocation system and the entities involved. In order to understand which are the measures that current key entities can apply to restrict the ability of third parties to communicate, we sketch the basic notions of the Internet routing system, and the cryptographic security standards proposed.\newline \indent
In the next section, we introduce the blockchain technology, and one of its incarnations, Ethereum, as it is the platform of choice for developing the InBlock solution.\newline
Section \ref{sec:InB} is devoted to present the rationale for the design of InBlock. For doing this, we analyse each of the requirements for an address allocation system, and how they are fulfilled by InBlock. Besides, Ethereum restrictions are analysed, to show that price, latency or throughput of blockchain operations are appropriate for address allocation needs.\newline \indent
In section \ref{sec:PE} we present an experiment in which InBlock is used to allocate a small set of addresses. We describe the objectives of the experiment, and how to run it in controlled, yet realistic, conditions. \newline \indent
We next describe related work, and we end with the conclusions.

The proposed approach falls between the policy and the computer science arenas. In this paper we focus on the policy side and we try to present the design of the InBlock from the perspective of the policy goals. While we try to keep the technical side to the minimum, we do need to describe some specific technical aspects to properly describe the proposed design.

\section{\label{sec:BackgroundIp}Background on IP address allocation and routing}

\subsection{Address allocations}
\begin{figure}[h!]
	\centering
	\includegraphics[width=1\linewidth]{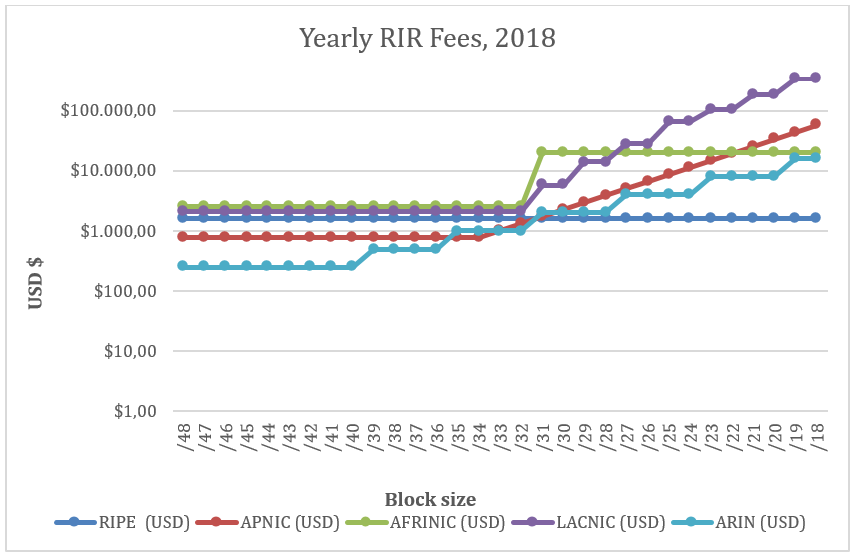}
	\caption{Yearly RIR Fees for Provider Aggregatable address assignments, 2018}
	\label{fig:rir1}
\end{figure}

IP addresses identify the end-points of every Internet communication and serve for providing both identity and location functions. An IP address is a 32-bit identifier for IPv4, and a 128-bit identifier for IPv4’s intended replacement, IPv6. Each network is assigned one or many ranges of IP addresses (called prefixes), which are non-overlapping with the prefixes assigned to other networks. The administrator of each network then assigns IP addresses to the nodes at the network. The assignment process must consider the limitation of the pools at the time of allocation, and the need to ensure uniqueness and proper registration to meet several operational requirements \cite{HCHC}, \cite{Mueller10}. \newline \indent
In the early days of the Internet, the responsibility for assigning and managing the global pool of IP addresses was performed by the University of Southern California as part of a research project funded by the Defence Advanced Research Projects Agency (DARPA), U.S. Government. Between 1998 and 2016, the National Telecommunications and Information Administration (NTIA), part of the Department of Commerce, U.S. Government, outsourced the technical management role to the ICANN, the Internet Corporation for Assigned Names and Numbers. In 2016, Internet stakeholders, including the US government, agreed to let the contract with the U.S. government expire and ICANN continued to perform the management of the global pool of Internet addresses, numbers and names ever since \cite{ICANN16}. Although ICANN’s structure and governance seeks for accountability and transparency, the ICANN is ultimately subject to the jurisdiction of the California State, in which many lawsuits have been filed \cite{ICANN18}. \newline \indent
ICANN has delegated the address assignment duties to the five Regional Internet Registries, RIRs, namely AFRINIC (Africa), APNIC (Asia-Pacific region), ARIN (mainly US and Canada), LACNIC (Latin America and the Caribbean), RIPE (Europe, Middle East and Central Asia). RIRs are open membership-based bodies composed primarily of organizations that operate networks. The address resources received from ICANN are assigned according to policies developed regionally by each RIR, although coordinated with the rest of them. Then, the resources are allocated to their requesters, according to six goals explicitly agreed among all RIRs in its policies \cite{RIPE18},\cite{APNIC18},\cite{ARIN18},\cite{LACNIC18},\cite{AFRINIC18}:

\begin{itemize}
	\item	Uniqueness: addresses must be globally unique, the \textit{raison d´être} of the registry.
	
	\item 	Fairness: Current policies are designed to be fair in the sense that they should be equally applied to all parties irrespectively of “their location, nationality, size, or any other factor” \cite{RIPE18}. 
	
	\item	Conservation. A main goal of the Internet resource allocation policies is to make a rational use of them and avoid wasteful practices.
	
	\item	Aggregation: The core routers of the Internet networks exchange information about the Internet address space assigned to each network to perform the global routing function. In particular, these routers must store and process advertisements for the prefixes that describe the address space assigned to every network, so the advertisement of a route can be seen as an externality \cite{Mueller10}. The lower the number of prefixes exchanged, the lower the hardware requirements imposed to all the routers participating in the interdomain routing system. The allocation policies aim to reduce the number of prefixes advertised by fostering hierarchical allocation to some extent. In particular, allocation policies encourage the use of provider-based address aggregation as a preferred choice, by giving large address blocks of so-called Provider Aggregatable (PA) addresses to network providers, which in turn suballocate them to end users. In this way, the routes to many different end users can be advertised by a single announcement, thus reducing the number of entries to store and process in the core routers. RIRs may also perform Provider Independent (PI) assignments directly to end-users, usually smaller, although they may be unreachable due to network operators not assuming the cost of routing them \cite{RIPE-pi}.
	
	\item 	Registration: Contact and other information associated with allocations is stored to help the normal operation of the Internet, such as serving to troubleshoot connectivity incidents. The current IANA-RIR based system maintains both a private and a public database with information regarding the allocations: first, any registry allocating an address block keeps (internal) records on the party receiving the resources. This typically involves a contract which includes detailed contact information. This information is not public and it is not used for Internet operations. It can be used for legal purposes as long as the legal actions are valid within the legal context of the host country. In addition, the party obtaining the resources may use any of the available Internet Routing Registry (IRR) to publish contact and technical information related to the resources. The information stored in the IRR system is reported to be incomplete or inaccurate in many cases \cite{KM17}. 
	
	\item 	Minimised overhead: the allocation system should work with as little overhead as needed to fulfil its function.
	RIRs can allocate Provider Aggregatable (PA) blocks to Local Internet Registries (LIRs) and they also provide direct Provider Independent (PI) assignments directly to end-users. All RIRs define a minimum PA IPv6 allocation of /32 \cite{RIPE18},\cite{APNIC18},\cite{ARIN18},\cite{LACNIC18},\cite{AFRINIC18}. The allocations can be (much) larger if the applicant justifies the needs. In particular, the larger allocations so far are /20. The minimum PI allocation is a /48 or a /56 depending on the RIR and they can be larger if justified.\newline \indent
	RIRs charge a yearly membership fee to entities holding Internet resources. In all RIRs except for RIPE, fees vary according to the amount of resources received. A /48 PI allocation fee is in the range of a few hundreds of US\$ (between US\$ 100 and US\$ 800, depending on the RIR). A /32 PA allocation fee ranges between US\$ 1000 and US \$ 2,500. See figure \ref{fig:rir1} for the detailed information on the fees.\newline \indent
	The current fee structure used by the RIRs is not lineal with the number of addresses. The fee for a /32 is roughly one order of magnitude larger than the fee for a /48 while a /32 contains $2^{16}$ more addresses than a /48. A similar effect can be observed in PA allocations of different size, meaning that the fee for a /20 is significantly less than  $2^{12}$ times the fee for a /32.
	
\end{itemize}

\subsection{Interdomain routing}
The Border Gateway Protocol (BGP) is used to exchange prefix reachability information between the different networks in the Internet. The function the BGP protocol performs is called interdomain routing. The different networks participating in the BGP protocol are identified through AS numbers, 32-bit unique identifier. The AS numbers are managed in a similar way than IP addresses through ICANN and the RIRs. \newline \indent
The original BGP specification lacks of security features enabling the manipulation of routing information. For example, an attacker can advertise someone else’s prefix as its own, to hijack the traffic for that prefix. In order to prevent such incidents, the BGP ecosystem has been recently enhanced with origin validation capabilities.  In order to address these limitations, additional security features have been recently added to the interdomain routing architecture. \newline \indent
Origin validation is provided by the Resource Public Key Infrastructure (RPKI) \cite{LK12} architecture. The RPKI architecture defines a distributed repository that contains Route Origin Attestations (ROAs), X.509 certificates that are used to assert that a network (identified its AS number) is authorized to advertise a given prefix BGP. The trust chain starts from the RIRs, that issue certificate delegating prefix ranges to the LIRs, etc., which can in turn issue ROAs. The RIRs also provide access to repositories with the cryptographic information issued so far, which at the time of this writing, comprise 8738 ROAs \cite{MIRO}. ROAs can be used to discard BGP advertisements, and thus, prevent reachability to certain prefixes, AS numbers, or combinations of both. \newline \indent 
Currently, each RIR has its own RPKI root certificate i.e., each RIR has issued self-signed certificates for the whole address space range of IPv4 and IPv6 \cite{NRO17}. Therefore, a network willing to perform origin validation for the routes to the networks of any region of the world can configure the root certificates of the five RIRs as Trust Anchors, and download the repositories they hold. Then, they can use this information to build the router configuration that will filter BGP route information in real time. \newline \indent
\cite{HCRG} shows that resource certificate issuers can manipulate at will the contents of their publications; for example, an entity can issue certificates to invalidate or supersede previous resource delegations, affecting connectivity with or through networks performing origin validation. Top-level RPKI entities, the RIRs, can invalidate or supersede resource allocations they previously allocated, so they can prevent communication for any network with a certificate rooted in the Trust Anchor of the RIR considered. \newline \indent
Moreover, because each RIR has a root certificate for the whole IP address space, any RIR can issue valid certificates for any IP prefix. This means that a RIR can issue a certificate for prefix for which another (valid) certificate, rooted on a different RIR, exists. For example, although prefix \textit{P} was assigned by RIR\_A to an entity in its region, out of its own address pool, and a certificate rooted at RIR\_A for prefix \textit{P} exists, RIR\_B may issue a conflicting certificate for the same prefix. The resulting behaviour of these actions vary for different third-party networks and routers, and will depend on the configuration of the software applying the validation rules over the RPKI information. The network exposed to the conflicting certificates may decide to select one of the certificates (based on explicit preferences on the Trust Anchor, or in the order in which the information has been processed) or to discard both. The implications are that the isolated action of a RIR over a prefix can invalidate it for all the networks performing any type of cryptographic validation if the prefix belongs to its region, and can interfere in ways that are hard to predict for prefixes belonging to other RIRs. 

\section{\label{blockchainBack}Background on Blockchains}

A blockchain is an \textit{immutable distributed ledger that records validated transactions permanently without the need of a trusted third party}. \newline \indent
The blockchain is a distributed ledger because all the information is stored in all the nodes composing the blockchain peer-to-peer network \footnote{ This is different than other distributed databases, where different parts of the database are stored in different nodes and there could be some level of replication just for redundancy and performance benefits}. \newline \indent
The blockchain is composed of a growing list of blocks, securely linked between each other through cryptography \cite{ZXDCW}. Every block contains a hash pointer to a parent block, a timestamp and transactions´ data. \newline \indent
The addition of new valid block is determined through a distributed consensus mechanism. The consensus is an emerging artefact representing the agreement reached by over than thousands of nodes on the blocks added to the blockchain. The most popular consensus mechanism is Proof-of-Work (PoW) \cite{MXZXQ}.\newline \indent
In PoW, nodes try to solve a complex mathematical problem in order to gain the right to append a block to the existent chain (and make some profit). New block signers are chosen through a \textit{mining race}. Every time a block is added, a new mining race starts and every miner tries to find the solution to gain the possibility of mining a block and receive the related fee. Since every miner is working on the same problem, once it is solved by someone, the computational power spent from the others on the same problem is wasted. PoW consensus mechanism is expensive in terms of energy consumption. \newline \indent
The hash structure of the blockchain makes computationally unfeasible to alter the data of one block without the manipulation of all subsequent blocks. Tampering the ledger then requires both the collusion of the majority of the network and an enormous amount of computational power to rebuild the chain from the replaced block. This is the sense in which we interpret the \textit{immutability} of the blockchain.  \newline \indent
Finally, we stress that only valid transactions are included in the blocks forming the blockchain. To determine if a transaction is valid, all nodes participating comply with the same block validation rules. For example, a value transaction must be signed with the private key of the originator. \newline \indent
The blockchain paradigm can be extended to the automation of complex resource manipulation and transference procedures in a transparent and trustable manner, by means of the specification of smart contracts.\newline \indent
A \textit{smart contract} \cite{P17},\cite{YLYHTD} is a programme that is stored in the blockchain and is executed by the nodes of the blockchain network. Once deployed in the blockchain, the blockchain nodes will execute the smart contract whenever a monetary transaction to the contract account occurs. Then, every node validating the blockchain will execute the code of the contract, written in the blockchain itself, reaching the same final state.  \newline \indent
Ethereum \cite{B18} is a public blockchain platform created to facilitate the development of smart contracts. Ethereum has a built-in Turing-complete programming language that allow developers to easily write smart contracts. Every operation in the network is triggered by transactions between accounts, that can be externally owned accounts (EOAs), owned and controlled by users, and contract accounts, associated to a smart contract which code and state  stored with the account itself. Being a public blockchain, any party can create one or more EOAs and any party can run a smart contract in Ethereum. \newline \indent 
Ethereum has implemented a PoW-based consensus mechanism. Miners are rewarded in Ether, the Ethereum cryptocurrency, for the storage and processing power they contribute to. Users that want to run a smart contract or to use one issue a transaction in the Ethereum network which includes a transaction fee payable to the miners. The value (in Ether) of the transaction fee is set by the user generating the transaction and should reflect the number of operation steps to be performed to accomplish a certain work and the priority that the user wants to get from the blockchain miners, as higher transaction fees imply that the transaction will be processed earlier by the miners. Transaction confirmation times are estimated around 10 to 15 seconds, depending on storage needs, code complexity and bandwidth usage. \newline \indent
Ethereum enables the deployment of a \textit{Decentralized Autonomous Organization} (DAO) \cite{B_18_2}, an organization that is fully implemented in the form of one or more smart contracts without any human involved in the daily operation of the organization. In particular, the bylaws of the organization are embedded into the code of the smart contracts. DAO’s financial transaction records and program rules are maintained on a blockchain.

\section{\label{sec:InB}Description of the proposed InBlock IPv6 registry}

InBlock is a Decentralized Autonomous Organization that performs IPv6 address allocation registry functions. By autonomous, we mean that the whole organization lies in the blockchain, in the form of smart contracts that run in the blockchain without human intervention. It is decentralized because the smart contracts run on the nodes that are part of the blockchain. The behaviour of the registry is governed by the code of the smart contract. Modifications to the information regarding the registry of IPv6 address blocks are triggered by blockchain transactions and subject to the consensus mechanism of the blockchain. Therefore, the smart contracts define what a valid transaction is and then all the nodes of the blockchain will enforce that only valid transactions modify the address allocation registry information. Once included in the blockchain, the allocation of a block is irrevocable. \newline \indent
InBlock is configured with a block of globally routable IPv6 addresses to allocate. When an entity wants to obtain an address allocation, it uses its Ethereum account to perform a request. The request is basically a blockchain transaction that transfers a predetermined fee (paid in Ether, the Ethereum cryptocurrency) to InBlock. InBlock verifies that the transaction is valid and that the fee has been correctly transferred. Upon reception of the transaction, the InBlock code goes through its associated state (stored in the blockchain) and finds an address block that is not currently allocated. InBlock associates the available block with the identity of the entity performing the request. This allocation information is recorded in the blockchain, and is associated to a predefined lifetime. The holder of the resources can renew the allocation making a new transaction transferring the yearly fee to the InBlock before the expiration date. If this happens, InBlock extends the lifetime of the allocation for another period. \newline \indent
Each IPv6 allocation record stored in the blockchain contains the information about the allocated prefix, the holder Ethereum Identity, the expiration date and a pointer to additional information about the holder of the allocation, allowing the holder to include contact or other route policy information. \newline \indent
As stated above, InBlock defines the rules that govern the IP address allocation. Phrased in the Internet Registry jargon, this means that the smart contract will encode the IPv6 address \textit{allocation policy}. It is only natural then that the goals for the design of the InBlock are aligned with the goals of the existent IPv6 address allocation policies. We next describe the different mechanisms that are part of InBlock aimed to fulfil the address assignment goals of uniqueness, registration, aggregation, conservation, fairness and minimum overhead stated by the current RIR-based allocation system (see section \ref{sec:BackgroundIp}).

\subsection{Uniqueness}
To guarantee uniqueness in the address assignment, we first require that the block of globally routable IPv6 addresses assigned to InBlock is unique (reserved exclusively for this purpose). Then we rely on the InBlock code, stored state, and the blockchain consensus mechanism to ensure that only unique assignments are valid, and thus, included in the blockchain registry.
\subsection{Conservation}
We identify two different concerns affecting conservation, namely stockpiling prevention and the reclaim of unused addresses. We describe the mechanisms used by the InBlock to deal with each of them separately.
\subsubsection{Stockpiling prevention}

One major concern to be considered when designing an IP address registry is how to prevent stockpiling, i.e., the accumulation of resources beyond the actual legitimate needs of the requesting entity \cite{RIPE18}, \cite{APNIC18}, \cite{ARIN18}, \cite{AFRINIC18}. IP addresses are valuable assets and given the precedent regarding IPv4 address space exhaustion, some parties may be tempted to obtain IPv6 addresses as a precautionary measure against a possible future scarcity. The InBlock design must provide the means to prevent or at least to control the extent of stockpiling. The current IANA-RIR system de-facto uses four mechanisms to prevent IPv6 address stockpiling. First, they require a justification for the need of the resources requested (based on planned needs for IP addresses for initial allocations and based on the HD ratio metrics regarding subsequent allocations). Second, the requirement to become member of the RIR, which implies paying an initial fee, and the charge of a yearly fee to the parties holding resources, with the fee in most cases related to the amount of resources received (see \ref{fig:rir1}). Third, an abundance argument: there are enough addresses for all future needs, so there is no point in stockpiling. And fourth, the possibility of reclaiming the addresses allocated if the holder of the resources does not comply with the requirements defined in the allocation policy. \newline \indent
In InBlock, we explicitly give up the first and the fourth mechanisms. In order to achieve full de-centralized control, the process of granting a new allocation and the process of renewing it must be fully automatic and encoded in the blockchain. Both the first and the forth aforementioned mechanisms require some form of human intervention, making them incompatible with the InBlock design goals. We argue that the remaining two mechanisms, fees and abundance of addresses, are enough to prevent stockpiling in the IPv6 case. \newline \indent
We next need to define the fee and allocation size structure that suits the InBlock purposes and deters stockpiling efficiently. We use as a starting point the current size and fee structure used by the RIRs. As mentioned in section \ref{sec:BackgroundIp}, the current fee structure used by the RIRs is not lineal with the number of addresses.\newline \indent
This poses a challenge for a fee-based automated mechanism aimed to deter address waste such as the InBlock. The lower the cost per address, the less effective the mechanism to deter stockpiling and other wasteful practices is. On the other hand, if the fee is set to the largest cost per address currently used by the RIRs (e.g., to the cost per address used in /48 PI allocations), this would render the cost of a larger block impractically high (the cost of a /32 would be tens of millions of US\$ if the cost per address of a /48 is used).\newline \indent
It is challenging for the InBlock to have different cost per address depending on the size of the allocation, because this may incentivize applications for larger blocks even when not needed, resulting in address waste (note that we do not have a complementary mechanism such as a need assessment procedure, to modulate user requests). On the same vein, having a low cost per address in larger allocations may promote a secondary market, where it is possible to obtain larger allocation for the same fee (or less) that it would take to obtain a smaller allocation directly from the InBlock\footnote{ Note that this is a risk of the current fee system, in which the cost grows in a sublinear trend with the number of addresses.}. This would again create the incentives for applicants to obtain larger allocations than what they would really need through the secondary market, resulting yet again in address waste. \newline \indent

\begin{figure}[h!]
	\centering
	\includegraphics[width=1\linewidth]{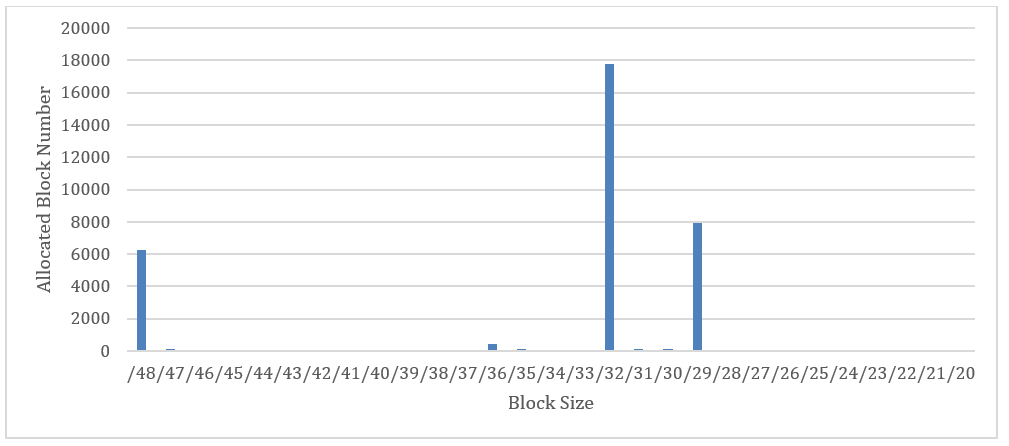}
	\caption{Distribution of number of blocks allocated up to May 2018, per block size.}
	\label{fig:rir2}
\end{figure}

In figure \ref{fig:rir2}, we depict the distribution of block sizes of all the existent allocations (for all the RIRs\footnote{ftp://ftp.afrinic.net/pub/stats/afrinic/delegated-afrinic-extended-latest 
	ftp://ftp.apnic.net/pub/stats/apnic/delegated-apnic-extended-latest\\
	ftp://ftp.arin.net/pub/stats/arin/delegated-arin-extended-latest\\
	ftp://ftp.lacnic.net/pub/stats/lacnic/delegated-lacnic-extended-latest\\
	ftp://ftp.ripe.net/pub/stats/ripencc/delegated-ripencc-extended-latest}).
We observe that, at the time of this writing, there are 17,795 allocations of /32, 6,283 allocations of /48, 7,903 allocations of /29. There are 191 allocations larger than /29. So, roughly half of the existent allocations are /32s, 25\% are /29s and the remaining 25\% are PI allocations of /48. \newline \indent
Considering all the above, we propose that the InBlock only allocates /32s and /48s, charging a fee slightly higher than the one set by the RIRs. By doing this, we can satisfy the most common allocation sizes. If an entity requires more than a /32, it probably can afford to request 2 or even 4 /32s, paying the corresponding fee for each of them. Since the fee increases linearly with the number of addresses for allocations larger than a /32, we believe that this fee structure would be enough to deter request additional /32 blocks that are not really needed. \newline \indent
This still does not address the problem regarding wasteful allocations involving allocations/assignments smaller than /32. Current RIRs fees are such that a /32 costs in the range of US\$ 2,500 to US\$ 1,000, and a /48 (PI) costs in the range of US\$ 100 and US\$ 800 (plus an additional initial fee in the range of US\$ 250 to US\$ 2,500). Suppose that the InBlock charges US\$ 3,000 for a /32 and US\$ 300 for a /48. Since the fee is one order to magnitude larger for a /32 than for a /48, this is likely to be enough to avoid the majority of applicants that would satisfy their needs with a /48 to request a /32, reducing address waste.\newline \indent
A final argument to support that the proposed fee structure is sufficient to deter stockpiling is the following: If the fee of a /32 is 3,000 US\$, then getting the whole IPv6 address space would imply a total amount of $12.6*10^{12}$ US\$ (compared to the world Gross Domestic Product, GDP  which is $ 76*10^{12} $ US\$) making it impossible for any party to even get hold of a significant chunk of the IPv6 address space.\newline \indent
Some further considerations about the fees.

\begin{itemize}
	\item	Currency: InBlock will define the fee in a fiat currency (Euros, US\$, Yuans) but the InBlock will actually collect the fee in the cryptocurrency used in the blockchain (Ether in Ethereum). Because the prices of the cryptocurrencies fluctuate significantly, we propose to define the fee in the fiat currency and convert the fee from the fiat currency to the current correspondence in Ether. 
	\item 	Fee Update. Fees cannot be constant because if the fiat currency value devaluates, the fee will less effective as a mechanism to deter stockpiling. On the other hand, InBlock is designed to work without human interaction. We propose to link the evolution of the fee to the increase of the world GDP. In this way, the fee will be updated to preserve its value in the future. Also, making the update of the automatic (and preventing any human interaction when defining the fees in the future) provides a stable framework for parties holding resources and prevents from having a human deciding on the fees (which may give the human the power to arbitrarily increase the fee, and to push the parties out of the system).
	
	\item	What to do with the fee money? It is worth to note that the fee, in the ranges discussed above, is just a mechanism to prevent stockpiling. The actual cost of running the InBlock is way much less than the fee (see below, we estimated 15 US\$). Several institutions involved in the resources (ICANN, RIRs, IETF, ISOC) could be natural recipients of this money, or the money could just be destroyed (sent to an nonexistent account). \indent
	
\end{itemize}

\subsubsection{Reclaiming unused addresses}
With the exhaustion of IPv4 addresses, attention has been paid to policies to reclaim unused addresses. However, only market-based approaches have proven to be effective, as coercive measures face many challenges \cite{Levin14}, \cite{Mueller10}. The InBlock expiration model provides a mechanism to neatly signal the Internet community which address blocks are in interest for its holder. When an address allocation is not renewed before its expiration date, the address block returns to the pool and the address space is eligible for further reassignment. \newline \indent
Besides, this mechanism serves to fix a well-known issue with storing tokens in blockchains: if the holder of the resource loses the private key that secures the allocation, the resource is lost. In particular, it is estimated that 20\% of existing Bitcoins are “lost” because of this \cite{FORT}.  

\subsection{Aggregation}
Another important consideration that allocation mechanism needs to address is related to the preservation of the global routing table. Due to the large size of the IPv6 address space, fostering aggregation is crucial for the viability of the routing system. Current RIR allocation policies promote aggregation through the preferred use of PA addresses, but they still allow PI allocations. PI allocations contribute to the size of the global routing tables as they cannot be aggregated. In the previous section we presented a fee structure to deter parties from requesting larger allocations than needed. In this section, we discuss the factors that hinder the deaggregation resulting from a secondary market of address blocks. We also present how InBlock enables aggregation of multiple address blocks obtained by the same entity through a sparse allocation policy. \newline \indent
\subsubsection{Deagggregation of address blocks due to secondary market transfers}
One potential concern with the proposed fee structure is that it may incentivize the creation of a secondary address market that may extend the use of PI allocations by end sites as opposed to PA allocation from the LIRs. \newline \indent
Currently RIRs charge a fee for each PI allocation. While the current policies do impose a number of requirements on end sites requesting PI block, these requirements are usually in the form of a plan or a prevision regarding the number of hosts/sites in the near future. We argue that because the InBlock charges a similar fee than the RIR for a PI /48 allocation, it is unlikely that there will be an increased demand of PI allocations due to the InBlock. \newline \indent
The current fee structure would make economically attractive for a party to obtain a /32 from the InBlock and re-sell smaller allocations cheaper than the corresponding fee from the InBlock. This may render PI allocations more affordable, and parties that would not be willing to pay they yearly fee that the RIR or the InBlock charges for a /48 allocations, may be willing to obtain a much cheaper /48 from this secondary market. Since these would indeed be PI allocations (as they are not provided by the ISP), they are likely to be announced in the global routing table as separated routes, bloating the routing table. Note that PI allocations are attractive because they avoid provider lock-in and the associated renumbering cost if the customer changes ISP. \newline \indent
However, a /48 obtained in the secondary market is not a perfect substitute of a /48 obtained directly from the InBlock. The reason is that in the /48 obtained through the secondary market, there is an intermediary, the reseller. If the intermediary fails to renew the /32 where the /48 are extracted from (i.e., because it goes bankrupt), all the /48 allocations will not be renewed neither and the end sites will lose their addresses. Moreover, once that the end site has obtained an allocation from the intermediary and configured in their network, the intermediary has an incentive to increase the charged fee, since there is a cost from the end user to renumber its network. The end site would then be trading \textit{ISP lock-in} for \textit{intermediary lock-in}. Nevertheless, it is still possible that some sites find it attractive to obtain PI blocks in the secondary market, and negatively impacting the global routing table. In order to prevent this, the InBlock could limit the number of different AS numbers used in ROAs within a single /32 to a maximum number (e.g., 100 different AS numbers). This restriction would not impose a real restriction in the operations of ISPs using the /32 for PA allocations, but would negatively affect the intermediaries that want to resell PI allocations out of a /32, since it would prevent all the end sites obtaining an allocation out of a single /32 to use different origin ASes in their ROAs (which is what they would naturally do when announcing a PI block in the interdomain routing).\\
\subsubsection{Sparse allocation}
In order for the multiple blocks assigned to a single entity to be aggregatable, the InBlock will use a sparse allocation strategy \cite{WPP} to manage the overall pools. This allows the holder of a resource to request for the contiguous block, so that the multiple blocks that a given entity obtains from the InBlock are aggregatable. \newline \indent
When submitting a new request for a new block, the applicant can attach a proof that it holds a block from a previous allocation. If this is the case, InBlock will allocate a contiguous block, enabling the aggregation of the two prefixes. 

\subsection{Registration}
In the case of the InBlock, blockchain identities are mostly anonymous\footnote{Identities are not perfectly anonymous, since there are means to try to link an Ethereum identity to a physical entity, but in general, it is not required prior identification to obtain a Ethereum identity.}. Moreover, payments are done using the corresponding Ether, which as of today provides significant anonymity features. All this implies that the InBlock has no information about the entity that received the allocation. InBlock provides the means to voluntarily include contact information or route policies for each allocation, in a similar way to Internet Routing Registries \cite{KM17}, but it does not mandate it. However, it is worth to note that the blockchain enables a cryptographic link between the holder of the account to which the address prefix has been assigned, and the routing information. This means that any other party accessing to this information can verify that the contact and routing policy information is authorized by the assignee of the Internet resources. \newline \indent
In summary, InBlock provides stronger privacy features than the current IANA-RIR system while still enabling resource holders to voluntarily provide contact and routing information for operational purposes. Besides, InBlock lacks strong traceability features available in the IANA-RIR system for lawful purposes. We argue that this is an appropriate approach and that the traceability for legal reasons should be pursued through the ISPs providing connectivity to the allocated address block. The reason why we find this appropriate is that the InBlock provides a global service, hence it should avoid the asymmetry existing in the current IANA-RIR system between the country hosting the registry and the rest of the countries. Pursuing legal actions through the ISPs providing connectivity to the allocated prefix seems to provide a better match between the scope of the organization and the scope of the legal framework.  

\subsection{Fairness}
Fairness is an explicit goal in current RIR address allocations policies. Fairness in this context means that the policies should be equally applied to all parties irrespectively of “their location, nationality, size, or any other factor” \cite{RIPE18}. InBlock naturally achieves that goal, since any party can obtain an Ethereum identity and then it can obtain an allocation from the InBlock. \newline \indent
Moreover, InBlock takes a step further in the fairness domain, and achieves \textit{jurisdictional fairness}. As stated earlier, one explicit goal of the InBlock design is to prevent the so-called jurisdictional overflow, where the allocations of entities in one country are under the jurisdiction of another country. We can phrase this in terms of fairness, i.e., that every entity has the right that its address allocations are not ruled by the legal framework of another country. This is not the case today, since allocations to entities based in the countries where the RIRs are based are ruled by their national law system, while for allocation to all other entities are ruled by foreign legal systems. 

\subsection{Minimised overhead}
By all accounts, the InBlock operation is very efficient and provides reduced overhead. As there are no humans involved in the operation, the costs are very low (about US\$ 15, see section \ref{sec:EthTech}). Also, the time-scale of operation of the InBlock is significantly shorter than the current system. Allocations in InBlock are completed in the order of minutes (see section \ref{sec:EthTech}).

\subsection{\label{sec:EthTech}Ethereum Technological aspects}

We now discuss technological aspects related to the specific platform of choice, Ethereum. We justify the selection of Ethereum for this purpose. Then we cursorily analyse if Ethereum can provide the latency, throughput and cost required by InBlock. \newline \indent
We design InBlock as a set of smart contracts on top of Ethereum. Previous proposals \cite{HL16},\cite{PREC},\cite{KCEBN} suggest to create a new blockchain to store information regarding IP address allocation (see the Related work section for further details about these proposals). In InBlock, instead of creating a new blockchain, we propose to use an existing one. We believe this approach provides three important benefits. \newline \indent
First, it provides a clean architecture with layered design that separates the blockchain from the registry service. This allows the evolution of the blockchain without affecting the registry service. 
%For example, there is an ongoing debate regarding whether Proof or Work approaches are sustainable due to their expensive cost in terms of power consumption and whether Proof of Stake provides a more sustainable alternative. 
By laying the InBlock on top of an existing (and evolving) blockchain, we make InBlock agnostic to the consensus mechanism. In particular, Ethereum currently uses PoW (which is the proven technology) and it is experimenting with PoS \cite{CASPER}, as it may be a more sustainable alternative. Once/if PoS is proven and stable, Ethereum will migrate to PoS and InBlock will benefit from this technological advance without any impact. \newline \indent
Second, by using an existing blockchain, we can rapidly develop and deploy InBlock. For example, Ethereum is already available and working. By using Ethereum, we reduce the development time, as we only need to focus in the implementation of the registry service. All blockchain code evolution and testing\footnote{Notable bugs have been identified in different blockchains in the past \cite{20}.}  are taken care of by the Ethereum community. \newline \indent
Third, using an existing blockchain provides secure bootstrapping, i.e., a secure blockchain from the start of InBlock deployment. Blockchain security heavily depends on the consensus mechanism used (e.g., PoW, PoS). The level of security of the consensus mechanism frequently depends on the level of adoption of the blockchain. A blockchain using PoW is as strong as the hashing power used to mine blocks \cite{A14},\cite{MXZXQ}. In a new blockchain, it is likely that there will be little hashing power as there is little economical gain from mining it. If PoS is used instead, the level of security depends on the number of stakeholders, the concentration of stake and also how much actual value is there in the blockchain \cite{POS_ETH},\cite{MXZXQ}. In a new PoS blockchain, it is likely that there will significant concentration of stake and a reduced number of stakeholders control the blockchain. In the case of InBlock, the IPv6 addresses managed by the InBlock registry are valuable per se, prior to the existence of the blockchain. So, if the blockchain provides little security in its early days, there is a risk of an attack directed towards the illegitimate acquisition of allocations. The InBlock must then provide strong security from the bootstrap to be viable and using Ethereum achieves this goal. \newline \indent
Due to the facilities provided to develop smart contracts, and its relative maturity, Ethereum is our blockchain of choice. While the use of Ethereum seems to provide many advantages when building InBlock, we need to verify that Ethereum is also a good fit in terms of performance and cost. We next analyse different relevant parameters:
\begin{itemize}
	\item Timescale: There are roughly three delays involved in an Ethereum transaction, namely, the time it takes for a miner to include a transaction in a block that it is mining, the time it takes to mine the block containing the transaction and the time it takes for the block containing the transaction to be confirmed. Regarding the first delay, miners receive many transactions that are candidates to be included in the next block to be mined. Miners determine which transactions may be included in the current block according to the transaction fee offered (transactions that are willing to pay more get in the blockchain earlier). According to current prices \cite{GAS}, if the InBlock transaction is willing to pay US\$ 2 per allocation, the delay to be included in the next block is less than 2 minutes. Regarding the second type of delay, Ethereum publishes a new block every 17 seconds. Finally, assuming that InBlock will wait for 12 new blocks to confirm the transactions containing the allocations \cite{TIMES}, this means that it will take about 3 more minutes to confirm the allocation transaction. So, the total delay for an InBlock transaction to be published and confirmed in Ethereum will be in the order of 5 minutes. Currently it takes days to grant new allocations, so the timescale provided by Ethereum is much shorter than the one provided by the current allocation system. 
	\item Throughput: Ethereum currently has a mean throughput of 20 transactions per second. In order to estimate the maximum throughput that could be required by InBlock, we compute the number of transaction that it would require renewing yearly all the current IPv4 and IPv6 allocations, plus the transaction resulting from the new IPv4 and IPv6 allocation done every year. This results in 58,700 transactions per year, 0.0019 transactions per second, which is much less than the transaction throughput currently supported by Ethereum.
	\item Cost. In order to run in Ethereum, InBlock needs to pay to the Ethereum network in the form of a transaction fee. Executing InBlock implies the following expenses. First, the deployment of the InBlock code in the Ethereum network requires a transaction fee. We estimate roughly between US\$ 400 and US\$ 800 for this. Assuming that the InBlock is managing a /20 address block, this results in less than US\$ 0.5 per block (but this cost needs to be paid upfront when InBlock is launched). Once deployed, we estimate in around US\$ 15 the costs of the transactions required to assign a block. There are other transactions required to run InBlock that will incur in additional costs (like calls to the different oracles, see below) but these can be considered negligible. Note that the cost of a transaction fee is likely to change with time, and it is expected that the it will increase as Ethereum becomes popular. If InBlock defines a fixed transaction fee according to current values, it is possible that this value will become outdated and that InBlock transaction become unattractive for miners to include them in new blocks they are mining. Because of this, the transaction fee used by the InBlock must be updated according to reflect the values expected by the miners at every time. This can be done in Ethereum implementing an “oracle” \cite{GAS_ORA} that provides the InBlock with updated values to use for transaction fees. 
	
\end{itemize}

\section{\label{sec:PE}Proposed experiment}

As the blockchain technology is fairly new, probably susceptible to bugs, and fast evolution, and the IP address allocation apparatus is a cornerstone of the Internet, we are not proposing to create an operational IPv6 registry using InBlock. Instead, we are promoting to do an experiment with a limited scope, to gain a better understanding of how blockchain technology can be used to provide registry services for Internet resources and moreover, if the notion of a distributed autonomous registry is feasible.\newline \indent
Through the proposed experiment, we believe we can gain a better understanding of the following aspects: 

\begin{itemize}
	\item	Whether the proposed approach based on a yearly fee is enough to deter stockpiling.
	\item 	Whether the addresses obtained from the InBlock are actually used and announced in the Internet. Whether the blocks are announced as they are assigned, or extra deaggregation is observed.
	\item	Whether a secondary market of addresses emerges.
	\item	Whether the number of PI allocations increases.
	\item 	Whether the contact information of the allocated blocks is completed and updated with sound contents.
	\item 	How changes in the Ethereum affect the InBlock service. In particular, when bugs appear, how much they affect the InBlock service and whether it is possible to cope with them. Also, how technological leaps in Ethereum affect the InBlock service (e.g., if Ethereum moves from PoW to PoS, how this affects InBlock).
\end{itemize}

The experiment, by definition, has a limited lifetime. The lifetime should be defined in advance, before starting the experiment. A reasonable timeframe could be five years. If the experiment is a success, the lifetime of the experiment can be extended and the allocations made renewable for as long as the holders want to renew them. If the experiment is terminated, the allocations made through the InBlock should be formalized upon its termination in one of the existent Internet registries. This would allow organizations participating in the InBlock experiment and using the allocated addresses to avoid the cost of renumbering regardless the result of the experiment. With this conditions, the InBlock experiment is likely to be more realistic, as more organizations may use the resources obtained through the InBlock in real operations. \newline \indent
The experiment should include some safeguards in case an unexpected behaviour is observed. For instance, it would be beneficial to rate limit the number of allocations made. If the rate in the applications is excessive, the experiments should be paused and pass the control to human supervisors. \newline \indent
In order to run the InBlock experiment, a pool of globally unique addresses is needed. The organizations that can provide the required pool of addresses and run the experiment are: the IETF (80\% of the IPv6 address space is “Reserved for the IETF” \cite{SPACE}, and it is up to the IETF to instruct IANA to allocate an address block for the InBlock experiment), ICANN through the IANA, the RIRs (either one RIR on its own, or jointly through the NRO), or a NIR (National Internet Registry). \newline \indent
In addition of providing the addresses required to perform the experiment, the organization running the experiment should also be capable of properly designing and implementing it. In this document we propose InBlock as a starting point for that debate, but in order for the experiment to be useful, the Internet community should be actively engaged in the ultimate design of the experiment. Similarly, once that the experiment has been defined, it needs to be implemented and tested. The resulting implementation should be reviewed by the community to make sure that there are no errors or backdoors. We believe that there are several organizations who have the capabilities to run the proposed experiment, such as the IETF or the RIRs. \newline \indent

\section{Related work}
We next describe existing work related to the management of Internet resources using blockchain technology. In particular, we attempt to identify how the previous work has addressed the challenges pointed out in section \ref{sec:InB}.\newline \indent
Namecoin \cite{KCEBN} is a working system that provides decentralized namespaces improving security, privacy and censorship resistance.  \newline \indent
Namecoin supports name resolution for several namespaces, including the DNS names within the .bit domain. While the .bit domain belongs to the globally DNS hierarchy, the mechanism for resolving .bit names is different than other DNS names, as it requires querying the Namecoin blockchain for name resolution. This implies changes in the host resolver. Technically, Namecoin is a fork of Bitcoin with a new native token, the Namecoin (NMC) token. As a consequence, Namecoin uses a PoW consensus mechanism. The main challenge of the use of PoW for a new coin is the lack of economic incentives for miners to mine. In order to address this issue, Namecoin uses merged-mining with Bitcoin. Namecoin DNS names are paid using NMC tokens. The price of the NMC token is set by the market. In theory, the NMC token price would be high enough to prevent stockpiling. In reality, there was an initial peak on the demand for attractive names such as well-known trademarks, then a drop to its current value that is close to US\$ 0. Analysis showed that most of the names initially acquired belong to three entities, and are not used for name resolution, so it should be assumed that were acquired for speculation. \newline \indent
As a take away, Namecoin is a valuable experience for designing blockchain based systems for managing Internet resources. First, the Namecoin experience shows that the fee mechanism must use a high-enough fee to be effective. Leaving the market to determine the fee seems unwise, especially in the early phase, when there is little demand, the market laws make the price to decrease, which in turn render the fee mechanism ineffective.  \newline \indent
However, there are fundamental differences between IP addresses and DNS names, namely: first, while all IP addresses have the same value, DNS names do not. Some DNS names are more attractive than other (e.g., those reflecting trademarks). Second, in order to use the .bit names allocated though Namecoin, host modifications are required, while no modification in end hosts are required to use IP addresses allocated through the InBlock. \newline \indent
Internet blockchain \cite{HL16} proposed the use of blockchains to register IP addresses. They consider a Bitcoin-like blockchain for the management of all Internet resources including IP addresses, ASes, DNS and BGP route. The main goal is to avoid centralization and the control of a single authority. To reduce design and development risks associated to the blockchain technology, they choose the Bitcoin as its underlying platform. They propose using multi-sig (a technology in which a cryptographic operation can be performed or validated with a minimum number of keys from a larger set) to alleviate the key-lost problem. The Internet Blockchain research concludes that this technology is still not mature enough to support BGP route advertisement. For such a reason they propose an incremental deployment, starting from the management of RPKI functionalities, then proceeding to support route advertisements, and finally providing name resolution (currently solved by the DNS system). The paper does not address key concerns such as stockpiling prevention or secure bootstrapping. \newline \indent
Palisse et. al. \cite{PREC} propose a blockchain mechanism to manage IP addresses. A new blockchain would record IP address allocation and delegation, based on a PoS consensus mechanism, where addresses are the main asset. The main concern regarding such approach is that in PoS blockchains, the ones having more stakes are the ones with more chances to mine new blocks. This means that the current authority of the global IPv6 address pool (IANA) would certainly control the majority of the stakes and probably the blockchain, defeating the goal of preventing centralized control of the IP address allocation system. \newline \indent
Like Namecoin \cite{KCEBN}, they propose an expiration mechanism with renewal to prevent the lost-keys problem and to preserve the consistency of the Internet resources. This work does not consider concerns such as stockpiling or secure bootstrapping.\newline \indent
Kuerbis and Mueller \cite{KM17} discuss the applicability of blockchain technology to perform the Internet Routing Registry functions. Unforgeability and a provable timeline are identified as desirable properties of a routing policy registry, and are naturally provided by blockchains. They consider a system in which operators could cryptographically point to routing policy information repositories by using Blockstack, a blockchain which facilitates decentralized naming systems. Integration with RPKI key data can ensure the authenticity and integrity of the data. InBlock is similar in the sense that it stores the minimal metadata in the blockchain, and points elsewhere to the storage of the information. In this case, the authenticity and integrity of the routing policy data is cryptographically linked to the identity to which the addresses where allocated.

\subsection{Alternative IP address allocation systems}
There have been proposals for alternative mechanisms to manage IP address allocations. In particular, there have been several proposals for geographically aggregatable addresses (the original IPv6 address architecture reserved an address range for geographic-based unicast addresses \cite{RFC1884}). The main concern with geographic aggregation is that it opposes to the business logic of the ISPs, since it would require that ISPs covering a given geographic area to announce the routes for the prefixes of that area and thus carry traffic for customers of the competing ISPs in the region. InBlock does not argue for any form of geographical aggregation. Addresses allocated by InBlock can be used as any other global Unicast Address assigned through the current IANA-RIR based system. \newline \indent
Transferable Block Lease (TABL) \cite{MK09} proposes a market-oriented IPv6 allocation mechanism. They propose that RIRs set aside an address block that can be allocated without the assessment of the needs to the applicant, with a fee that is related to the size of allocation. The block allocated though TABL would be transferable between parties. They argue that such an approach would provide a simpler and cheaper method for address management and would create the incentives for resource holders to return the unused address blocks. TABL is similar to the InBlock proposal in the sense that they both propose a fee based mechanism to acquire IP address allocations and in the analysis of how the resulting mechanism would work. The main difference is that the TABL still relies in the IANA RIR hierarchy for the whole address management process, so the IANA and the RIRs still control the address allocation (revoking and/or preventing parties to obtain addresses) while the use of blockchain technology in the InBlock limits the power of the allocation hierarchy. 

\section{Conclusions and final remarks}
The current IP allocation system exhibits the so-called jurisdictional overflow problem, namely, that the consequences of the application of the legal framework of the 5 countries where the RIRs and IANA are based overflows its geographical boundaries, \textit{de facto} making entities in other countries subject to their legal system in matters related to address allocation. This may reasonably cause concern for some countries and entities within those countries. We believe that blockchains provide the means to build alternative approaches to the global management of IP addresses that do not exhibit the aforementioned jurisdictional overflow problem. \newline \indent
There are many different ways that blockchains can be used to build a global IP address registry. We described InBlock, a decentralized autonomous organization running in Ethereum that performs the IPv6 registry functions. Its design represents a specific set of trade-offs. In particular, InBlock implements a registry that is not under control of any single entity, implying that no single entity can prevent another entity to obtain IP address allocations.  Moreover, InBlock provides quite strong privacy guarantees and censorship resistance. Nevertheless, as a trade-off, InBlock gives up the traceability and enforcement features. This means that the IPv6 address registry functions provided by InBlock cannot be used to trace the identity of the holders of the resources if they do not voluntarily include contact information in the registry. Similarly, the InBlock cannot be used to restrict the Internet access to any entity by revoking their address allocations (actually it is a goal of the InBlock to prevent this situation). \newline \indent
The proposed InBlock design makes a particular set of tradeoffs. Other tradeoffs are also possible and would result in alternative approaches with different characteristics. As future work we plan to work in a permissioned version of the InBlock. A permissioned InBlock would include an additional entity in the architecture, the \textit{InBlock registrars}. The InBlock registrars would be responsible for validating the identities of the entities applying for resources. Once an entity obtains a valid InBlock identity from the registrar, it can apply for IP address allocation from the permsssioned InBlock in a similar fashion than in the described InBlock. The role of registrars could be fulfilled by the current RIRs, NIRs, or new organizations with national/regional scope. Compared to InBlock, this mechanism provides much improved traceability. This traceability is achieved at the cost of requiring the permission of one of the organizations controlling the blockchain. However, compared to the current allocation system, no hierarchy would exist, and the capacity to revoke allocations could be largely restricted. Therefore, even the perisssioned InBlock would provide a better match the existing jurisdictional scope than the current system, and thus, could represent a better fit between the Internet and the geopolitical governance. \newline \indent
The InBlock design described here aims to achieve some defined goals (uniqueness, stockpiling prevention, aggregation preservation, etc.) by means of some mechanism (blockchain consensus, pricing, limits on the number of AS number involved in the delegations, etc.). The extent to which these goals can be achieved with these measures can only be derived from real experience. We also appreciate a lack of experience in the use of blockchains for solving real-world problems. Because of this, this is a call to arms for experiments on the use of blockchain technology for Internet governance in general and for implementing IP registry functions in particular. We propose to experiment with InBlock for IPv6 address allocation. We believe this experiment will provide useful hand-on experience about how blockchains can be used for managing allocations. Other experiments using other mechanisms that make different trade-offs would also be useful. These experiments would not only inform how to evolve Internet governance towards more decentralized approaches, but would also help to shape the future blockchain regulation.

%\section*{References}

%	\section*{Acknowledgements}
%	This work has been partially funded in the Huawei Innovation Research Prize (HIRP) InBlock.

\bibliographystyle{IEEEtran}

\end{document}